\documentclass[showpacs,prl,10pt,twocolumn,superscriptaddress,aps]{revtex4-1}

\usepackage{ae}
\usepackage{bm} 
\usepackage{xcolor}
\usepackage{amsmath}
\usepackage{amssymb}
\usepackage{amsfonts}
\usepackage{graphicx}
\usepackage{slashed}
\usepackage{wasysym}
\usepackage{upgreek}

\newcommand{\Eqref}[1]{Eq.~\eqref{#1}}

\makeatletter
\def\fps@figure{ht}
\def\fps@table{ht}
\makeatother

\allowdisplaybreaks

\usepackage{ulem}

\begin{document}

\setlength{\unitlength}{1mm}

\title{Enhancing quantum vacuum signatures with tailored laser beams}

\author{Felix Karbstein}\email{felix.karbstein@uni-jena.de}
\affiliation{Helmholtz-Institut Jena, Fr\"obelstieg 3, 07743 Jena, Germany}
\affiliation{Theoretisch-Physikalisches Institut, Abbe Center of Photonics, Friedrich-Schiller-Universit\"at Jena, Max-Wien-Platz 1, 07743 Jena, Germany}
\author{Elena A. Mosman}\email{mosmanea@tpu.ru}
\affiliation{National Research Tomsk Polytechnic University, Lenin Ave. 30, 634050 Tomsk, Russia}

\date{\today}

\begin{abstract}
 We demonstrate that tailored laser beams provide a powerful means to make quantum vacuum signatures in strong electromagnetic fields accessible in experiment.
 Typical scenarios aiming at the detection of quantum vacuum nonlinearities at the high-intensity frontier envision the collision of focused laser pulses.
 The effective interaction of the driving fields mediated by vacuum fluctuations gives rise to signal photons 
 encoding the signature of quantum vacuum nonlinearity.
 Isolating a small number of signal photons from the large background of the driving laser photons poses a major experimental challenge.
 The main idea of the present work is to modify the far-field properties of a driving laser beam to exhibit a field-free hole in its center, thereby allowing for an essentially background free measurement of the signal scattered in the forward direction. Our explicit construction makes use of a peculiar far-field/focus duality.
\end{abstract}

\maketitle

\paragraph{Introduction}

Maxwell's classical theory of electrodynamics provides an accurate theoretical description of the physics of macroscopic electromagnetic fields.
One of its cornerstones is the superposition principle, implying that light rays pass through each other without interaction and do not change their properties.
However, as a purely classical theory, Maxwell's electrodynamics should arise from the more fundamental theory of quantum electrodynamics (QED) in the formal limit of $\hbar\to0$.
The true theory of vacuum electrodynamics differs from Maxwell theory by terms suppressed parametrically with $\hbar$.
The leading corrections were determined as early as in the 1930s by Heisenberg and Euler \cite{Euler:1935zz,Heisenberg:1935qt}, who studied the effective self-interactions of slowly varying electromagnetic fields induced by vacuum fluctuations of electrons and positrons.
These non-linear couplings facilitate light-by-light scattering phenomena, which -- at least in principle -- invalidate the superposition principle.
Having no tree-level analogue, they are suppressed with inverse powers of the electron mass $m_e$, making them very elusive in experiment; see the Reviews~\cite{Fradkin:1991zq,Dittrich:2000zu,Marklund:2006my,DiPiazza:2011tq,Dunne:2012vv,Battesti:2012hf,King:2015tba,Inada:2017lop,Karbstein:2019oej} and references therein.

The advent of high-intensity laser facilities \cite{CILEX,CoReLS,ELI,SG-II} opens up the possibility of verifying QED vacuum nonlinearities in macroscopically controlled fields.
All-optical signatures, i.e., effects driven by laser fields resulting in photonic signals, are the prime candidates for such discovery experiments.
However, the large background of the driving laser photons typically constitutes a major obstacle in measuring the signal.
In this letter, we demonstrate analytically that tailored laser beams featuring a peak in the focus where the interaction takes place, but a hole in the far field where the signal is measured provide a novel means to overcome these limitations.
However, in order to achieve this we first of all need a self-consistent, analytic description of such laser beams. Throughout this work, we adopt Heaviside-Lorentz units with $c=\hbar=1$.

\paragraph{Pulsed laser fields}

For this endeavor, we only consider a specific class of monochromatic laser fields of oscillation frequency $\omega=\frac{2\pi}{\lambda}$, which are rotationally symmetric about the beam axis and are well-described by focused beam solutions of the paraxial wave equation: namely, linearly polarized laser fields without topological charge. These can be expressed as superposition of Laguerre-Gaussian ${\rm LG}_{p,l}$ modes \cite{Siegman,SalehTeich} with $l=0$, but finite $p\in\mathbb{N}_0$.
The paraxial approximation is valid for beams made up of photons with wave vectors $\vec{k}$ fulfilling $\vartheta=\varangle(\vec{k},\hat{\vec{\kappa}})\ll1$, where $\hat{\vec{\kappa}}$ is the direction of the optical axis of the beam.
To describe laser pulses of finite duration $\tau$, we supplement these beam solutions with a Gaussian pulse envelope.
This {\it ad hoc} prescription inevitably violates the wave equation at ${\cal O}(\frac{1}{\tau\omega})$.
For typical high-intensity (free-electron) laser pulse parameters $\tau\geq20\,{\rm fs}$ ($2\,{\rm fs}$) and
$\omega\geq1.5\,{\rm eV}$ ($3\,{\rm keV}$) we have $\tau\omega\gtrsim 43.2$ ($8700$), rendering this approximation well-justified.

Subsequently, we identify the beam axis with the $\rm z$ axis and use cylindrical coordinates $(r,\varphi,{\rm z})$.
For a linearly polarized laser pulse of the above type propagating in positive $\rm z$ direction we then have $\vec{E}(x)=\hat{\vec{e}}_EE(x)$, $\vec{B}(x)=(\hat{\vec{e}}_{\rm z}\times\hat{\vec{e}}_E)E(x)$ and $E(x)=\sum_p{\cal E}_p(x)$. The field profile associated with the ${\rm LG}_{p,0}$ mode focused at ${\rm z}=0$ reads
\begin{equation}
 {\cal E}_p(x)={\mathfrak E}_{p}\,{\rm e}^{-(\frac{{\rm z}-t}{\tau/2})^2}\,\frac{w_0}{w(\zeta)}\, L_p(2\chi^2)\,{\rm e}^{-\chi^2} \cos\bigl(\Phi_{p}(x)\bigr)\,, \label{eq:E}
\end{equation} 
with $\zeta=\frac{{\rm z}}{{\rm z}_R}$, $\chi=\frac{r}{w(\zeta)}$ and phase
\begin{equation}
    \Phi_{p}(x)=
    \omega({\rm z}-t)+\zeta\chi^2-(2p+1)\arctan\zeta+\varphi_p\,. \label{eq:Phi}
\end{equation}
Here, ${\rm z}_R=\frac{\pi w_0^2}{\lambda}$ and $w(\zeta)=w_0\sqrt{1+\zeta^2}$ are the Rayleigh range and radius of the fundamental Gaussian (${\rm FG}_0$) mode of waist $w_0$, and $\varphi_p$ is a mode-specific constant phase.
The relation between the peak field amplitude and the mode energy $W_p$ is $\mathfrak{E}_p^2\simeq8\sqrt{\frac{2}{\pi}}\frac{W_p}{\pi w_0^2\tau}$ \cite{Karbstein:2017jgh}.
We emphasize that apart from $\lambda$ and $\tau$, the field profile of a generic monochromatic laser field with zero topological charge is fully characterized by the set of $w_0$, $W_p$ and $\varphi_p$.

The laser intensity is defined as $I=|\langle \vec{E}\times\vec{B}\rangle_t|=\langle E^2\rangle_t=\sum_{p,p'}\langle{\cal E}_p{\cal E}_{p'}\rangle_t$, where $\langle\cdot\rangle_t$ denotes averaging over one laser period.
Up to subleading corrections of ${\cal O}(\frac{1}{\tau\omega})$, the pulse envelope is not affected by the averaging procedure.
This yields
\begin{align}
 I(x)&=4\sqrt{\frac{2}{\pi}}\,{\rm e}^{-2(\frac{{\rm z}-t}{\tau/2})^2}\frac{1}{1+\zeta^2}\sum_{p,p'}\frac{\sqrt{W_pW_{p'}}}{\pi w_0^2\tau} \nonumber\\
&\quad\times\cos\bigl[2(p'-p)\arctan\zeta+\varphi_p-\varphi_{p'}\bigr] \nonumber\\
&\quad\times L_p(2\chi^2)L_{p'}(2\chi^2)\, {\rm e}^{-2\chi^2}\,. \label{eq:I}
\end{align}
The pulse energy is $W=2\pi\int_0^\infty\!{\rm d}r r \!\int_{-\infty}^\infty{\rm d}t I(x)=\sum_p\! W_p$.

In the far field $\zeta\gg1$ or equivalently $\zeta\to\infty$, we have $\chi|_{\zeta\to\infty}\simeq\frac{r}{w_0\zeta}=\frac{\vartheta}{\theta}$, where $\vartheta$ is the polar angle measured relative to the beam axis and $\theta\simeq\frac{w_0}{{\rm z}_R}$ is the asymptotic beam divergence of the ${\rm FG}_0$ mode. The approximation $\vartheta\simeq\frac{r}{{\rm z}}$ is well-justified here as $\vartheta\ll1$ for paraxial beams by definition.
Hence, the far-field angular decay of the number of laser photons $N\simeq\frac{W}{\omega}$ with $\vartheta$ is $\frac{{\rm d}N}{\vartheta {\rm d}\vartheta}\simeq\frac{2\pi}{\omega}(\frac{w_0}{\theta})^2\int_{-\infty}^\infty{\rm d}t\,\zeta^2 I(x)|_{\zeta\to\infty}$.

Note, that in the focus at $\zeta=0$ the cosine in \Eqref{eq:I} becomes $\cos(\varphi_p-\varphi_{p'})$, while in the far field at $\zeta\to\infty$ it equals $(-1)^{p'-p}\cos(\varphi_p-\varphi_{p'})$.
This implies that only for $p'=p$, i.e., laser fields prepared in a pure ${\rm LG}$ mode, the radial intensity profiles in the focus and the far field have the same shape.
For generic laser fields, terms with $p'\neq p$ result in differently shaped focus and far-field profiles.

\paragraph{Flattened-Gaussian field profiles}
Following Refs.~\cite{Gori:1994,Karbstein:2017jgh}, it can be shown that in order to implement the flattened-Gaussian (${\rm FG}_{\cal N}$) radial field profile
\begin{equation}
 E_{\cal N}(\chi)\sim {\rm e}^{-\chi^2}\sum_{n=0}^{\cal N}\frac{1}{n!}\chi^{2n}\,,
 \label{eq:EN}
\end{equation}
with ${\cal N}\in\mathbb{N}_0$, at a given value of the longitudinal coordinate ${\rm z}={\rm z}_0$ ($\zeta_0=\frac{{\rm z}_0}{{\rm z}_R}$), the mode-specific phases and energies in Eqs.~\eqref{eq:E}-\eqref{eq:I} have to be chosen as
\begin{align}
 \varphi_p\ \to\ p(\pi+2\arctan\zeta_0)\,, \quad
 W_p\ \to\ \Bigl(\frac{c_{p,{\cal N}}}{C_{\cal N}}\Bigr)^2 W\,, \label{eq:subsFGB}
\end{align}
and the sums over $p$, $p'$ be restricted to all integers from $0$ to ${\cal N}$.
The coefficients in \Eqref{eq:subsFGB} are determined by the expansion coefficients of the Exponential Sum Function in the Laguerre basis, $\sum_{n=0}^{\cal N}\frac{1}{n!}\chi^{2n}=\sum_{p=0}^{\cal N}(-1)^p c_{p,{\cal N}}L_p(2\chi^2)$, and read \cite{Gori:1994,Karbstein:2017jgh}
\begin{equation}
 c_{p,{\cal N}}=\sum_{k=p}^{\cal N}\binom{k}{p}\frac{1}{2^k}\,, \quad C_{\cal N}^2=\sum_{p=0}^{\cal N}c_{p,{\cal N}}^2\,. \label{eq:cs}
\end{equation}
Note, that in \Eqref{eq:subsFGB} we effectively traded the mode-specific parameters $\varphi_p$, $W_p$ characterizing a more generic laser field for a subset parameterized by ${\cal N}$ and $\zeta_0$.
We denote the field profile of the laser pulse featuring a ${\rm FG}_{\cal N}$ radial profile~\eqref{eq:EN} at $\zeta=\zeta_0$ by $E^{\zeta_0}_{\cal N}(x)$.
Generically, the larger $\cal N$, the wider and the more flat-top-like the ${\rm FG}_{\cal N}$ radial field profile~\eqref{eq:EN} for fixed $\zeta_0$ and $w_0$ \cite{Gori:1994,Karbstein:2017jgh}.

Notably, these findings also allow us to implement laser fields exhibiting a field-free hole in the center of their transverse profile at $\zeta=\zeta_0$.
To achieve this, we subtract two ${\rm FG}_{\cal N}$ field profiles characterized by the same values of $w_0$ and $\tau$ but different $\cal N$ and $W$, i.e., determine $E^{\zeta_0}_{{\cal N,N'}}(x)=E^{\zeta_0}_{{\cal N}}(x)-E^{\zeta_0}_{{\cal N}'}(x)$ with ${\cal N}>{\cal N}'$, while ensuring both fields to have the same peak field amplitude at $\zeta=\zeta_0$.
The latter condition can be enforced by choosing the energy of the ${\rm FG}_{{\cal N}'}$ beam as $W' = (\frac{C_{{\cal N}'}}{C_{\cal N}})^2 W$.
The resulting expression for the ${\rm FG}^\circledcirc_{{\cal N},{\cal N}'}$ field closely resembles a single ${\rm FG}_{\cal N}$ field:
to implement it, the mode-specific phases and energies have to be chosen as in \Eqref{eq:subsFGB}, with the minor modification that now
\begin{align}
 c_{p,{\cal N}} \ \to\ c_{p,{\cal N}}-\Theta({\cal N}'-p)c_{p,{\cal N}'}\,, \label{eq:subsFGBdiff}
\end{align}
with Heaviside function $\Theta(\cdot)$.
The pulse energy $W_{{\cal N},{\cal N}'}$ is obtained by summing up the energies~\eqref{eq:subsFGB} with \Eqref{eq:subsFGBdiff}.
Field-free rings can be implemented along the same lines but require the superposition of more beams.

Finally, some general comments are in order.
From \Eqref{eq:subsFGB} it is obvious that for the special case of a flattened-Gaussian ${\rm FG}\in\{{\rm FG}_{\cal N},{\rm FG}^\circledcirc_{{\cal N},{\cal N}'}\}$ field profile to be implemented in the focus at $\zeta_0=0$ we have $\varphi_p\ \to\ p\pi$, while for a ${\rm FG}$ beam profile in the far field at $\zeta_0\to\infty$, we have $\varphi_p\ \to\ 2p\pi$.
This difference can be attributed to the additional factor of $(-1)^{p'-p}$ in the transverse intensity profile in the far field not present in the focus; cf. the 2nd paragraph below \Eqref{eq:I}.
Noteworthily, it implies the following far-field/focus dualities, which hold up to an obvious overall normalization factor: (i) The far-field intensity profile $I^{0}(x)|_{\zeta\to\infty}$ of a beam featuring a ${\rm FG}$ focus profile matches the focus profile $I^{\infty}(x)|_{\zeta=0}$ of a beam with a ${\rm FG}$ far-field profile and vice versa.
(ii) In the two special cases characterized by either $\zeta_0=0$ or $\zeta_0\to\infty$ the far-field $I^{\zeta_0}(x)|_{\zeta\to\infty}$ and focus $I^{\zeta_0}(x)|_{\zeta=0}$ profiles are related by the substitution $c_{p,{\cal N}}\to (-1)^p c_{p,{\cal N}}$.

\paragraph{Flattened-Gaussian far-field profiles}

\begin{figure}
 \centering
 \includegraphics[width=1\linewidth]{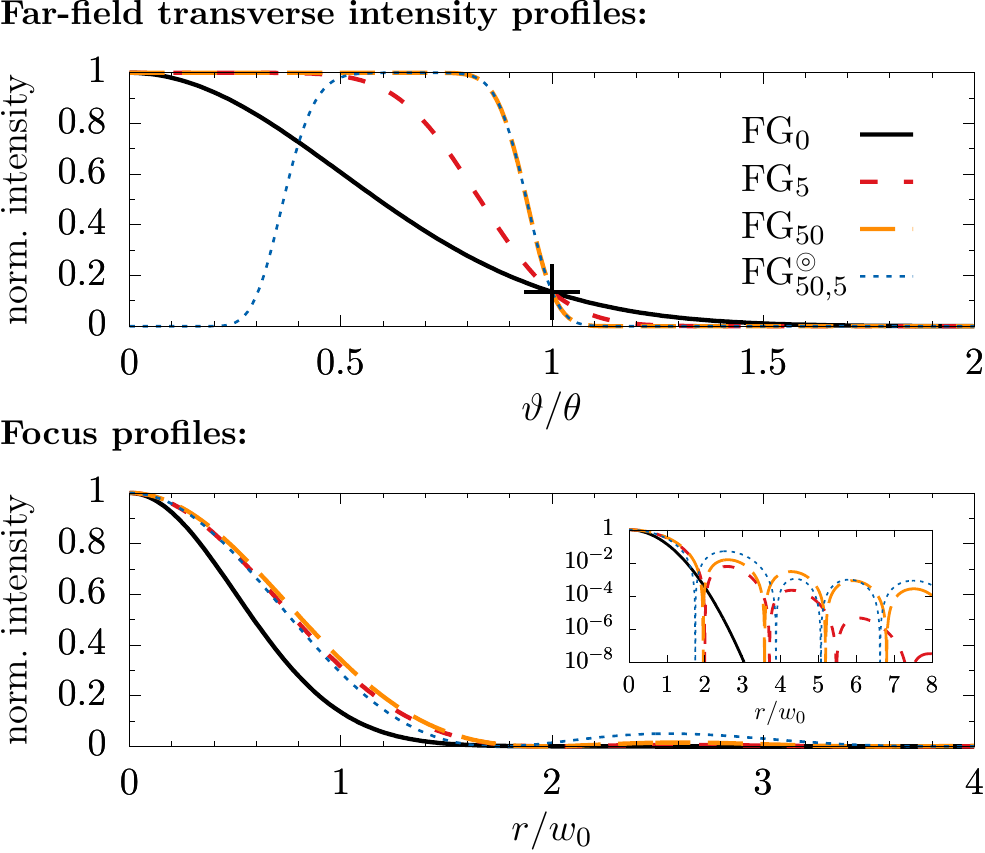}
 \caption{Upper panel: Far-field intensity profiles of laser fields with ${\rm FG}\in\{{\rm FG}_{\cal N},{\rm FG}^\circledcirc_{{\cal N},{\cal N}'}\}$ profiles implemented in the far-field as a function of $\vartheta$ measured in units of the asymptotic divergence $\theta$ of the ${\rm FG}_0$ beam. Here, the waist parameters $w_0$ determining the angular spread of the far-field profiles via $\theta\simeq\frac{\lambda}{\pi w_0}$ are rescaled such that $\theta_{\cal N}\to\theta$. The cross highlights the matching point.
 Lower panel: Associated focus profiles as a function of the beam radius measured in units of the ${\rm FG}_0$ beam waist $w_0$; the inset shows the same curves, adopting a logarithmic scale.
 Deviations between the curves and the emergence of Airy rings are particularly visible in the logarithmic plot.}
 \label{fig:Profiles}
\end{figure}

In the remainder of this letter, we focus on the special case of $\zeta_0\to\infty$ and thus only consider laser beams characterized by ${\rm FG}$ far-field profiles.
For a graphical representation of a selection of far-field and focus profiles of ${\rm FG}$ profiles implemented in the far field, see Fig.~\ref{fig:Profiles}.

To allow for a better comparison of the focus profiles associated with different ${\rm FG}_{\cal N}$ far-field profiles, we introduce effective waists $w_{\cal N}$. 
The latter are defined as the $1/{\rm e}^2$ radii of the intensity profiles $I^{\infty}_{\cal N}(x)|_{\zeta=0}$ in the focus, resulting in the defining equation $\sum_{p=0}^{\cal N} c_{p,{\cal N}}L_p \bigl(2s_{\cal N}^2\bigr)={\rm e}^{s_{\cal N}^2-1}\sum_{p=0}^{\cal N} c_{p,{\cal N}}$,
where the scaling parameter $s_{\cal N}=\frac{w_{\cal N}}{w_0}$ measures $w_{\cal N}$ in units of $w_0$.
For the present scenario, where the ${\rm FG}_{\cal N}$ profile is implemented in the far field, the widening of the ${\rm FG}_{\cal N}$ profile with $\cal N$ implies an increase of the effective beam divergence. This translates into  $w_{{\cal N}+1}< w_{\cal N}$, and thus $s_{\cal N}<1$ for ${\cal N}\geq1$.
Expanding the above defining equation up to quadratic order in $s_{\cal N}<1$, we obtain the estimate
\begin{equation}
 s_{\cal N}^2\,\approx\,\frac{(1-{\rm e}^{-1})\sum_{p=0}^{\cal N}c_{p,{\cal N}}}{\sum_{p=0}^{\cal N}(2p+{\rm e}^{-1})c_{p,{\cal N}}}\quad\text{for}\quad{\cal N}\geq1\,. \label{eq:scale}
\end{equation}
Similarly, the peak intensity in the focus can be related as $I_{\cal N}^{\infty}(x)|_{\vec{x}=0}=\sigma_{\cal N} I_0(x)|_{\vec{x}=0}$ to the focus peak intensity $I_0(x)|_{\vec{x}=0}$ of the ${\rm FG}_0$ laser field.
This scaling factor reads
\begin{equation}
 \sigma_{\cal N}=\frac{1}{C_{\cal N}^2}\Bigl(\sum_{p=0}^{\cal N} c_{p,{\cal N}} \Bigr)^2\,. \label{eq:peakI}
\end{equation}

Upon insertion of \Eqref{eq:cs}, the sums in Eqs.~\eqref{eq:scale} and \eqref{eq:peakI} can be performed, resulting in $s_{\cal N}^2\approx{\frac{2(1-{\rm e}^{-1})}{{\cal N}+2{\rm e}^{-1}}}$ and $\sigma_{\cal N}=(\frac{{\cal N}+1}{C_{\cal N}})^2$.
The analogous results  $w_{{\cal N},{\cal N}'}= s_{{\cal N},{\cal N}'}w_0$ and $I_{{\cal N},{\cal N}'}^{\infty}(x)|_{\vec{x}=0}= \sigma_{{\cal N},{\cal N}'}I_0(x)|_{\vec{x}=0}$ for a beam with an ${\rm FG}^\circledcirc_{{\cal N},{\cal N}'}$ far-field profile follow with \Eqref{eq:subsFGBdiff}, yielding
$s_{{\cal N},{\cal N}'}^2\approx\frac{2(1-{\rm e}^{-1})}{{\cal N}+{\cal N}'+1+2{\rm e}^{-1}}$ and 
$\sigma_{{\cal N},{\cal N}'}=(\frac{{\cal N}-{\cal N}'}{C_{\cal N}})^2$.
As the two ${\rm FG}_{\cal N}$ beams superimposed to form the latter fulfill ${\cal N}>{\cal N}'$ and $w_{\cal N}\leq w_0$, we have $s_{{\cal N},{\cal N}'}<1$.

Along the same lines, effective asymptotic divergences $\theta_{\cal N}$ can be defined as the polar angle for which the ${\rm FG}_{\cal N}$ far-field intensity drops by a factor of $1/{\rm e}^2$ relatively to its on-axis peak value. This results in the defining equation
$\sum_{n=0}^{\cal N}\frac{1}{n!}(\frac{\theta_{\cal N}}{\theta})^{2n}={\rm e}^{(\frac{\theta_{\cal N}}{\theta})^2-1}$, which is approximately solved by $\theta_{\cal N}\approx\theta\sqrt{1+{\cal N}}$.
Besides, the peak intensity in the far field scales as  $I^\infty_{\cal N}(x)|_{\zeta\to\infty}=C_{\cal N}^{-2}I_0(x)|_{\zeta\to\infty}$.
The inner and outer asymptotic divergences of a laser field featuring a ${\rm FG}_{{\cal N},{\cal N}'}^\circledcirc$ far-field profile are determined by $\theta_{\cal N}$ and $\theta_{{\cal N}'}$. Moreover, by construction the far-field peak intensity of such a beam equals $I^\infty_{\cal N}(x)|_{\zeta\to\infty}$.

Taking into account the large-$\cal N$ scaling of $C_{\cal N}^2\sim{\cal N}$ \cite{Gori:1994}, it can be easily verified that the above scalings consistently imply that for large values of $\cal N$ the product of the peak intensity and the beam radius squared is independent of $\cal N$.

\paragraph{Exemplary results}

Subsequently we demonstrate the novel possibilities enabled by the use of beams featuring a peak in the focus but a hole in the far field for nonlinear QED experiments on the example of vacuum birefringence \cite{Toll:1952rq}: linearly polarized probe photons traversing a strong pump field can pick up an ellipticity if their polarization vector has a non-vanishing overlap with the two distinct polarization eigenmodes imprinted on the quantum vacuum by the pump field. This results in polarization-flipped signal photons constituting the experimental signature of the effect.
The head-on collision of an XFEL probe and a high-intensity pump constitutes a promising route to its first measurement \cite{Aleksandrov:1985,Heinzl:2006xc,DiPiazza:2006pr,Dinu:2013gaa,Dinu:2014tsa,Karbstein:2015xra,Schlenvoigt:2016,Karbstein:2016lby,Karbstein:2018omb,Karbstein:2019bhp}.
Recently, it has been shown that employing the scattering of signal photons outside the forward-cone of the probe notably enhances the perspectives of measuring the effect for given parameters \cite{Karbstein:2015xra,Karbstein:2016lby,Karbstein:2018omb}.
Replicating this scenario with probe beams featuring a hole in the far-field seems particularly promising. In fact, it should facilitate an essentially background-free measurement of the signal photons scattered in the direction of the forward beam axis in a way not possible with conventional beams: while the focal spot of the probe photon field essentially does not differ from that of conventional beams (cf. Fig.~\ref{fig:Profiles}) and is in particular also characterized by a vanishing wave-front curvature, the driving probe photons are effectively diverted in the far field, leaving a field-free hole about the beam axis. On the other hand, the kinematics of the signal photons is determined by local properties of the driving fields in the interaction region about the beam focus, implying that the scattering phenomenon does not differ much from that induced by conventional beams. This results in signal photons quasi-elastically scattered in the direction of the beam axis. However, the hole in the far-field imprinted in the probe beam now allows for their unobstructed measurement.

All-optical quantum vacuum signatures are efficiently studied in the vacuum emission picture \cite{Galtsov:1971xm,Karbstein:2014fva}, encoding photonic signals of quantum vacuum nonlinearities in signal photons emitted from the strong-field region where the driving fields overlap.
The results presented below are based on the leading term of the Heisenberg-Euler Lagrangian~\cite{Euler:1935zz,Heisenberg:1935qt,Schwinger:1951nm}, allowing for the accurate theoretical study of all-optical QED vacuum phenomena driven by state-of-the-art XFEL and high-intensity laser fields.

Here, we consider the head-on collision of a generic ${\rm LG}$ probe with $l=0$ propagating in positive $\rm z$ direction with a ${\rm FG}_0$ pump (energy $\tilde{W}$, duration $\tilde{\tau}$, waist $\tilde{w}_0$ and Rayleigh range $\tilde{\rm z}_R$) at zero impact parameter. Both pulses are linearly polarized and the angle between their polarization vectors is $\frac{\pi}{4}\,{\rm rad}$, such that the number of signal photons scattered in a $\perp$-polarized mode is maximized.
As x-rays fulfill $\zeta\ll1$ throughout the interaction region, we can adopt the approximation $\zeta\approx0$ and thus $w(\zeta)\approx w_0$ when determining the vacuum emission signal \cite{Karbstein:2016lby}.
The intensity profile of the pump follows from \Eqref{eq:I} upon limitation to $p=p'=0$, substitution of ${\rm z}\to-{\rm z}$ and transition to tilded quantities.

Apart from the choice of the field profiles, this scenario matches the one detailed in Sec.~4.2 of Ref.~\cite{Karbstein:2019oej}: the differential number of $\perp$-polarized x-ray photons amounts to the $+$ component of Eq.~(44) with $p\ \to\ \perp$, while neglecting the manifestly inelastic terms in Eq.~(41) is equivalent to replacing the square of the pump field with its intensity profile.
Hence, the differential number of $\perp$-polarized x-ray photons of energy ${\rm k}=|\vec{k}|$ reads
\begin{align}
 &{\rm d}^3N_\perp\simeq\frac{{\rm d}^3k}{(2\pi)^3}\,\frac{32}{225}\alpha^4\frac{\rm k}{m_e^8}\Bigl|\sum_{p}{\cal M}_p\Bigr|^2\,, \label{eq:dNperp} \\
 \text{with}&\quad{\cal M}_p=\int{\rm d}^4x\,{\rm e}^{{\rm i}(\vec{k}\vec{x}-{\rm k}t)}\,{\cal E}_p(x)\big|_{\zeta\ll1}\,\tilde{I}(x)\,. \label{eq:Mpdef}
\end{align}
Accounting only for the dominant quasi-elastic contributions \cite{Karbstein:2015xra,Karbstein:2016lby}, \Eqref{eq:Mpdef} can be expressed compactly as
\begin{align}
 {\cal M}_p&\simeq\Bigl(\frac{8}{\pi}\Bigr)^{\frac{3}{4}}\frac{\tilde{W}}{w_0}\frac{\sqrt{\tau}}{T_+}\, {\rm e}^{-(\frac{\tau\tilde{\tau}}{4T_+})^2 ({\rm k}-\omega)^2}\sqrt{W_p}\,{\rm e}^{-{\rm i}\varphi_p} \nonumber\\
 &\quad\times\int_{-\infty}^\infty{\rm dz}\,\frac{[2-r^2({\rm z})]^p}{[2+r^2({\rm z})]^{p+1}}\,L_p\bigl(\tfrac{w_0^2k_\perp^2r^4({\rm z})}{2[r^4({\rm z})-4]}\bigr) \nonumber\\
 &\quad\times {\rm e}^{-\frac{(w_0^2k_\perp^2r^2({\rm z})}{4[2+r^2({\rm z})]}-2(\frac{4{\rm z}}{T_+})^2}\,{\rm e}^{{\rm i}\bigl[({\rm k}-\omega)(\frac{T_-}{T_+})^2+k_{\rm z}-\omega\bigr]{\rm z}} \,, \label{eq:Mp}
\end{align}
where $k_\perp^2=k_{\rm x}^2+k_{\rm y}^2$, $T_\pm=\sqrt{2\tau^2\pm\tilde{\tau}^2}$ and $r({\rm z})=\frac{\tilde{w_0}}{w_0}\sqrt{1+(\frac{\rm z}{\tilde{\rm z}_R})^2}$.
In the derivation of \Eqref{eq:Mp} we made use of standard algebraic operations and \cite{wolfram:id1}.
Subsequently we use spherical coordinates $k_{\rm z}={\rm k}\cos\vartheta$, $k_\perp^2={\rm k}^2\sin^2\vartheta$ and ${\rm d}^3k={\rm k}^2{\rm dk}\,{\rm d}\!\cos\vartheta\,{\rm d}\varphi$, and study the birefringence phenomenon with an ${\rm FG}^\circledcirc_{{\cal N},{\cal N'}}$ probe beam.
To this end we substitute \Eqref{eq:subsFGB} with coefficients~\eqref{eq:subsFGBdiff} into \Eqref{eq:Mp}.
\begin{figure}
 \centering
 \includegraphics[width=0.9\linewidth]{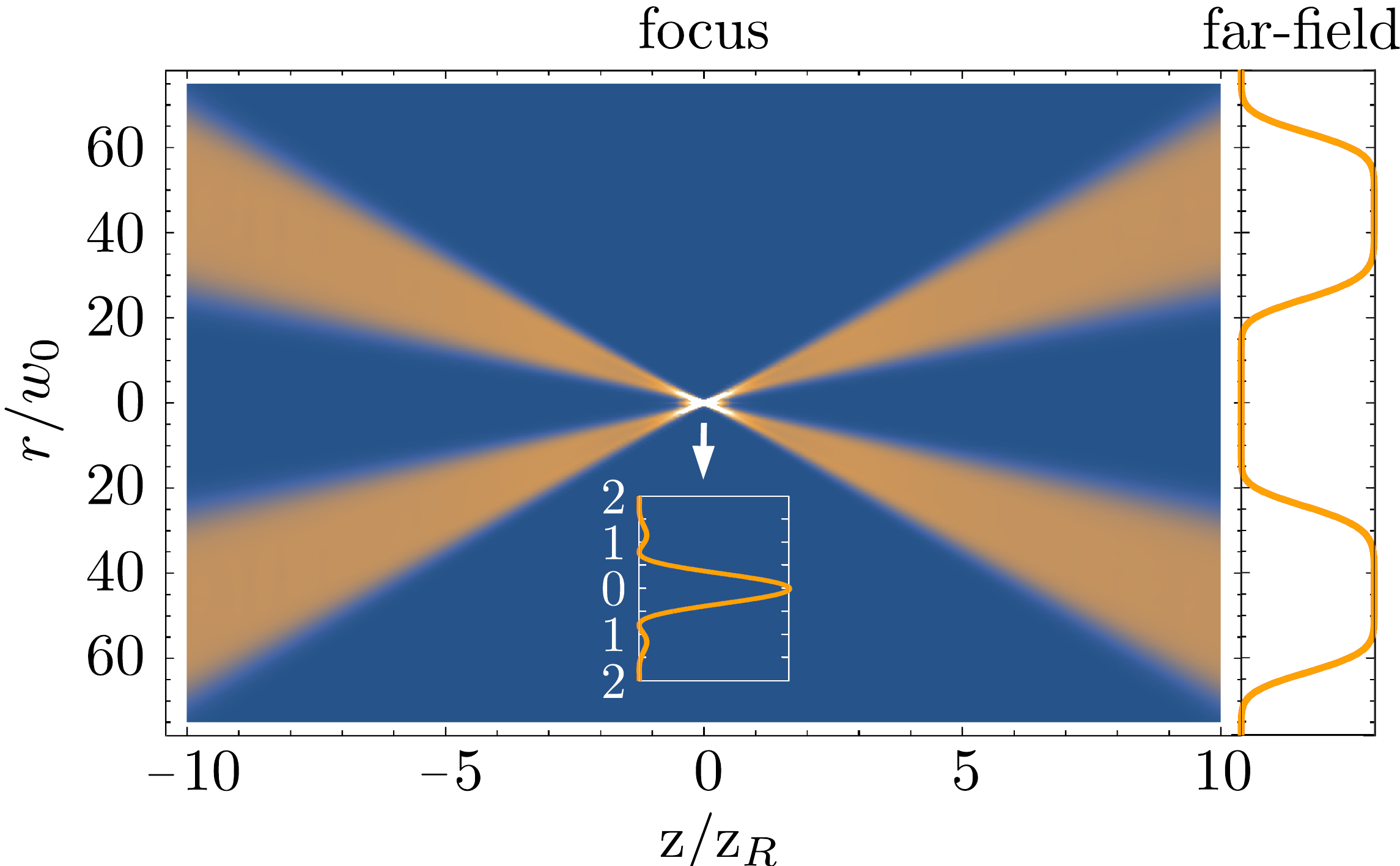}
 \caption{Laser intensity $I^{\infty}_{50,5}(x)$ as function of the longitudinal coordinate $\rm z$ measured in units of the Rayleigh range ${\rm z}_R$, and the radial distance $r$ measured in units of the waist $w_0$ of the ${\rm FG}_0$ mode.
 The color scale is such that the higher the intensity, the brighter the color: the intensity vanishes in the dark regions and reaches its maximum in the beam focus at $\vec{x}=0$.
 The inset shows the transverse focus profile; on the right we depict the far-field intensity profile.}
 \label{fig:BeamWithHole}
\end{figure}

\begin{figure}
 \centering
 \includegraphics[width=1\linewidth]{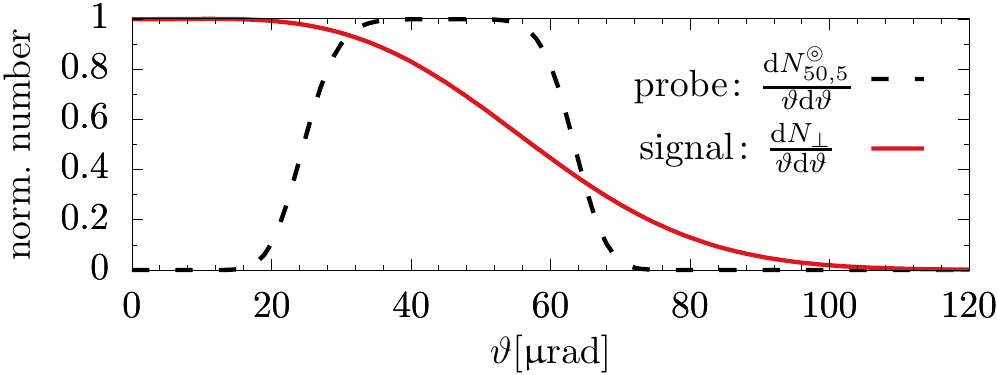}
 \caption{Far-field angular decay of the probe photons and the polarization-flipped signal photons for the parameters detailed in the main text.
 This figure highlights the potential of ${\rm FG}_{{\cal N},{\cal N}'}^\circledcirc$ laser fields in achieving an essentially background-free measurement of the signal at both small and large $\vartheta$, where the differential number of probe photons traversing the interaction region unaffected decays faster than the signal.}
 \label{fig:decay}
\end{figure}

In our explicit proof-of-principle example we assume the original XFEL probe to feature a ${\rm FG}_{50}$ far-field profile and deliver pulses of duration $\tau=25\,{\rm fs}$ encompassing $N_0=10^{12}$ photons of energy $\omega=12914\,{\rm eV}$; the pulse energy is $W=N_0\omega$.
The polarization of x-ray photons of this energy can be measured with a purity of ${\cal P}=5.7\times10^{-10}$ \cite{Marx:2013xwa}.
We envision the ${\rm FG}^\circledcirc_{50,5}$ profile to be obtained therefrom by adequately blocking the x-ray photons at small $\vartheta$, thereby reducing the number $N$ of photons in the ${\rm FG}^\circledcirc_{50,5}$ pulse; cf. Fig.~\ref{fig:Profiles} (upper panel) and Fig.~\ref{fig:BeamWithHole}.
Our considerations can be generalized to the specific, experimentally available laser fields acounting for the details of the blocking mechanism with the numerical code \cite{Blinne:2018nbd}.
For the pump we adopt the parameters of the high-intensity laser installed at the Helmholtz International Beamline for Extreme Fields (HIBEF) \cite{HIBEF} at the European XFEL \cite{XFEL}.
It delivers pulses of energy $\tilde{W}=10\,{\rm J}$ and duration $\tilde{\tau}=25\,{\rm fs}$ at a wavelength of $\tilde{\lambda}=800\,{\rm nm}$ and a repetition rate of $5\,{\rm Hz}$.
Here, we assume it to be focused to a waist of $\tilde{w}_0=1\,\upmu{\rm m}$. For the ${\rm FG}^\circledcirc_{50,5}$ probe we choose $w_0=3.3\,\upmu{\rm m}$, such that $w_{50,5}\simeq0.55\,\upmu{\rm m}$.
For these parameters, \Eqref{eq:dNperp} predicts $N_\perp\simeq1.47$ $\perp$-polarized signal photons per shot, to be compared with the total number of x-ray photons available for probing $N\simeq8.63\times10^{11}$; $N_0-N\simeq1.37\times10^{11}$ photons otherwise filling the hole are blocked.
Figure~\ref{fig:decay} shows that $\frac{{\rm d}N_\perp}{\vartheta\,{\rm d}\vartheta}$ reaches its maximum in a plateau at small values of $\vartheta\lesssim20\,\upmu{\rm rad}$ and slowly decays towards larger $\vartheta$. In the same angular interval, the photon distribution of the probe exhibits a hole, suggesting the possibility of an efficient signal-to-background separation.
Reducing $w_0$ for fixed $\tilde w_0$, the plateau disappears and the signal also starts to diminish towards small $\vartheta$, resulting in a peak at finite $\vartheta$.
Conversely, when increasing $w_0$ the plateau becomes narrower, eventually resulting in a narrow peak at $\vartheta=0$. Besides, the signal drops as less probe photons traverse the strong field region.
In turn, both variations tend to reduce the attainable signal.

Accounting for the finite purity of polarization filtering, only signal photons fulfilling $\frac{{\rm d}N_\perp}{\vartheta\,{\rm d}\vartheta}\geq{\cal P}\frac{{\rm d}N}{\vartheta\,{\rm d}\vartheta}$ 
can be measured above the background of the XFEL photons traversing the strong field region unaffected.
This criterion is met for $0\leq\vartheta\leq15\,\upmu{\rm rad}$ and $\vartheta\geq76\,\upmu{\rm rad}$.
The numbers of discernible signal photons per shot are $N_\perp^{\rm dis.}|_{\vartheta\leq15\,\upmu{\rm rad}}\simeq0.094$ and $N^{\rm dis.}_\perp|_{\vartheta\geq76\,\upmu{\rm rad}}\simeq0.14$.
While the former are emitted into a solid angle of $\approx700\,\upmu{\rm rad}^2$, the emission angle for the latter is $\approx30$ times larger.
The first signal is particularly promising. Its measurement only requires a detector with small angular acceptance and no recollimation is needed for an efficient detection.

\paragraph{Conclusions and Outlook}

In this letter we have constructed a new class of beam solutions to the paraxial wave equation featuring a field-free hole in its transverse profile at a given longitudinal coordinate. 
Using the nonlinear QED signature of vacuum birefringence as an illustrative example, we have demonstrated the great potential of beams featuring a peak in the focus, but a hole in the far field for strong-field QED experiments.
Certainly, many other signatures of quantum vacuum nonlinearity can be critically enhanced by such tailored beams.
This is especially true for scenarios based on the collision of high-intensity laser pulses, usually characterized by a paradox: while most signal photons arise from quasi-elastic scattering processes, for standard beams their measurement is typical obstructed by the background of the driving laser photons.
Thus, though containing much less photons, inelastic channels constitute the most prospective experimental signatures \cite{Karbstein:2019dxo}. 
Our proof-of-principle study underpins that the use of tailored beams can change this and make quasi-elastically scattered signals experimentally accessible.

\begin{acknowledgments}

This work has been funded by the Deutsche Forschungsgemeinschaft (DFG) under Grant No. 416607684 within the Research Unit FOR2783/1 and was supported by the German Academic Exchange Service (DAAD) under the ``Mikhail Lomonosov'' Programme.
Moreover, E.A.M. would like to thank the Helmholtz Institute Jena for hospitality and support. We are grateful to Matt Zepf for helpful discussions, and to Holger Gies for valuable comments on the manuscript.

\end{acknowledgments}

\end{document}